# The interplay between spatial and heliconical bond order in twist-bend nematic materials

R. Saha[3], C. Feng[1,2], C. Welch[4], G. H. Mehl[4], J. Feng[2], C. Zhu[2], J. Gleeson[3], S. Sprunt[1,3] and A. Jákli[1,3*]

[1]Chemical Physics Interdisciplinary Program, Advanced Materials and Liquid Crystal Institute, Kent State University, Kent, OH 44242, USA
[2]Advanced Light Source, Lawrence Berkeley National Lab, Berkeley, CA 94720, USA
[3]Department of Physics, Kent State University, Kent, OH 44242, USA
[4]Department of Chemistry, University of Hull, Hull, UK

*Abstract:*

*The nanostructure of two novel sulfur containing dimer materials has been investigated experimentally by hard and by resonant tender X-ray scattering techniques. On cooling the dimers through the nematic to twist-bend nematic (N-$N_{TB}$) phase transition, the correlation length associated with short-range positional order drops, while the heliconical orientational order becomes more correlated. The heliconical pitch shows a stronger temperature dependence near the N-$N_{TB}$ transition than observed in previously studied dimers, such as the CBnCB series of compounds. We explain both this strong variation and the dependence of the heliconical pitch on the length of the spacer connecting the monomer units by taking into account a temperature dependent molecular bend and intermolecular overlap. and. The heliconical structure is observed even in the upper 3-4°C range of the smectic phase that forms just below the $N_{TB}$ state. The coexistence of smectic layering and the heliconical order indicates a $SmC_{TB}$ -type phase where the rigid units of the dimers are tilted with respect to the layer normal in order to accommodate the bent conformation of the dimers, but the tilt direction rotates along the heliconical axis. This is potentially similar to the $SmC_{TB}$ phase reported by Abberley et al (Nat. Commun. **2018**, 9, 228) below a SmA phase.*

## *1. Introduction*

The formation of helical structure in molecular systems usually requires crystal or liquid crystal phases with chiral components. The smallest helical pitch formed by organic molecules is the α-helix of proteins[1] with *p=0.55 nm*, meaning *3.6* amino acids in L-configuration make one turn.[1] Such a tight pitch requires internal hydrogen bonding between chiral amino acid residues that join together in peptide chains that crystallize into a structure with long-range positional order. The helical pitch of chiral nematic liquid crystals (3-D anisotropic fluids) of rod-shaped molecules ranges from 0.1 to several hundred micrometers, i.e., hundreds to thousands of chiral molecules are needed to make one turn.[2] Recently achiral liquid crystal oligomers (dimers [3–19], trimers[20,21] and tetramers[12,21,22]) with rigid arms connected by odd-numbered methylene or ether groups were found to exhibit a twist-bend nematic phase ($N_{TB}$) [23–25], which features a heliconical structure with nanoscale pitch, p=6-20 nm range[6,26]. Remarkably such a helical structure requires neither hydrogen bonding, nor molecular chirality or positional order.

The first proof of the nanoscale pitch was provided by freeze fracture transmission electron microscopy (FF-TEM) measurements.[6,26] As they require rapid quenching of samples from well-defined different temperatures, detailed temperature dependence measurements would require large number of samples. The temperature dependence of the orientation (bond) ordering using only one sample was first probed with resonant soft X-ray scattering at carbon K-edge (RSoXS) [27–30]. Although RSoXS can be employed for all materials that contain carbon atoms, the λ=4.4 nm wavelength of the soft x-ray limits the resolution to a few nanometers and its *l*~0.3μm penetration depth requires the preparation of submicron thick films.[31] For dimer materials containing sulfur atom(s) Tender Resonant X-ray Scattering (TReXS) at the sulfur K-edges (E=2.471keV, λ~0.5nm, *l~10μm*) offers a more attractive alternative for precise measurement of the temperature dependence of the helical pitch of the $N_{TB}$ phase, as shown by several recent studies.[32–37]

In spite of the fact that small angle hard X-ray scattering is not suitable for pitch measurements due to the lack of an electron density modulation coupled to the heliconical structure, recent studies showed that careful analysis of synchrotron SAXS results can provide important information about the molecular associations both in the $N_{TB}$ phase and the N phase above it.[22] Combining SAXS and TReXS measurements allows one to probe the relation between short-range positional and longer range orientational order in the $N_{TB}$ phase.

In this paper we combine SAXS and TReXS measurements on two novel sulfur-containing analogues of fluorinated dimers with *n*-pentyl ($C_5H_{11}$) terminal chains.[38,39] In addition to the temperature dependences of the periodicities of the molecular associations and of the heliconical pitch, we also measure correlation lengths of the positional order of the molecular associations and of the heliconical orientational order. We find that below the N-$N_{TB}$ transition the positional order decreases to about 6 nm, while the bond-order increases to about 60 nm. These results imply that the molecular overlap increases at the onset of the $N_{TB}$ phase and lead us to propose a refined packing model of the heliconical structure that can explain both the reduction in positional correlations and the temperature dependence of the heliconical pitch. We also show that there is an about 4°C temperature range below the NTB phase where the smectic order and the heliconical order coexist and discuss its possible origin.

## 2. *Results and discussion*

### A. *Materials*

The synthesis towards the investigated materials **A** (Butyl(4"-(7-(2',3'-difluoro-4"-pentyl-[1,1':4',1"-terphenyl]-4-yl)heptyl)-2',3'-difluoro-[1,1':4',1"-terphenyl]-4-yl)sulfane) and **B** (butyl(4"-(11-(2',3'-difluoro-4"-pentyl-[1,1':4',1"-terphenyl]-4-yl)undecyl)-2',3'-difluoro-[1,1':4',1"-terphenyl]-4-yl)sulfane) is shown in Figure 1. The molecules **A** and **B** are characterized by two aromatic difluoroterphenyl units separated either by a heptyl or an undecyl spacer. The aromatic units are terminated on one side either by a pentyl group linked directly to the aromatic core or a butylthioether function, introducing the functionality for TReXS measurements.[31] The architectures are as close as possible to *DTCnCm* dimers reported earlier.[38,39] Notable is that for those materials it was not possible to obtain RSoXS results, probably due to the large number of different carbon atoms in the flexible terminal chains. The synthesis of **A** and **B** follows routes described previously and recently reported in detail by Stevenson et al.[40] New synthetic steps are associated with the formation of 4'-(butylthio)-2,3-difluoro-[1,1'-biphenyl]-4-yl)boronic acid (**2**), obtained from butyl(2',3'-difluoro-[1,1'-biphenyl]-4-yl)sulfane by a reaction with *n*-Butyllithium at -78°C under an $N_2$ atmosphere followed by an addition of trimethyl borate and subsequent acidification at room temperature by addition of 10% HCl. **2** was reacted in transition metal catalyst ($Pd(PPH_3)_4$) mediated coupling reactions with either **1a** or **1b** to obtain the target materials **A** and **B** in yields after purification of respectively 71% and 68%.

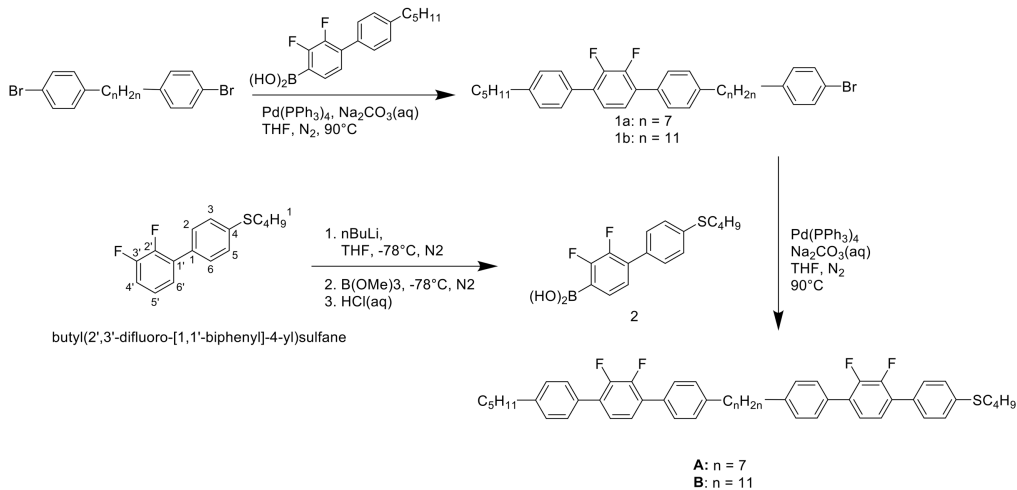

*Figure 1: Scheme of the Synthesis of the investigated systems **A** and **B**.*

Details of the synthesis are provided in the Supporting Information (SI).

The phase transition temperatures in °C as determined by DSC measurements on heating and cooling with *10°C/deg* rate are as follows. **A:** Cr 130.5 (SmX 107.5 $N_{TB}$ 122.5) N 144.8 Iso; **B**: Cr 90.5 SmX 109.3 $N_{TB}$ 125.0 N 161.7 Iso. Phases in parenthesis are monotropic, i.e., appear only on cooling from the isotropic phase.

### B. *Polarized Optical Microscopy (POM)*

Figure 2 shows the polarized optical microscopy (POM) textures of 5μm cells with planar alignment after cooling from the isotropic phase with *1 °C/min* rate. Figure 2 (a-d) and (e-h) show the textures of **A** and **B**, respectively. The uniform textures (a and e) represent aligned uniaxial nematic phase at 132 °C for **A** and 135 °C for **B**. Stripes parallel to the rubbing direction in Figure 2(b and f) at 118 °C for **A** and 116 °C for **B** are characteristic of the twist-bend nematic phase. Figure 2 (c) and (g) at 109.3 °C for **A** and 107 °C for **B** show the texture during (**A**) and 0.5°C below (**B**) phase transition to a smectic phase, respectively. During the transition twisted rope-like dendrites grow with the widths of the ropes and their final direction corresponding to the stripes of the $N_{TB}$ phase, as seen in Figure 2(c). After the transition being completed, the texture is characterized by twisted stripes running parallel to the stripes observed in the $N_{TB}$ phase. About 4-5 °C below the transition the stripes fade away, as seen in Figs. 2(d) and (h) at 104°C and 102°C for A and B, respectively.

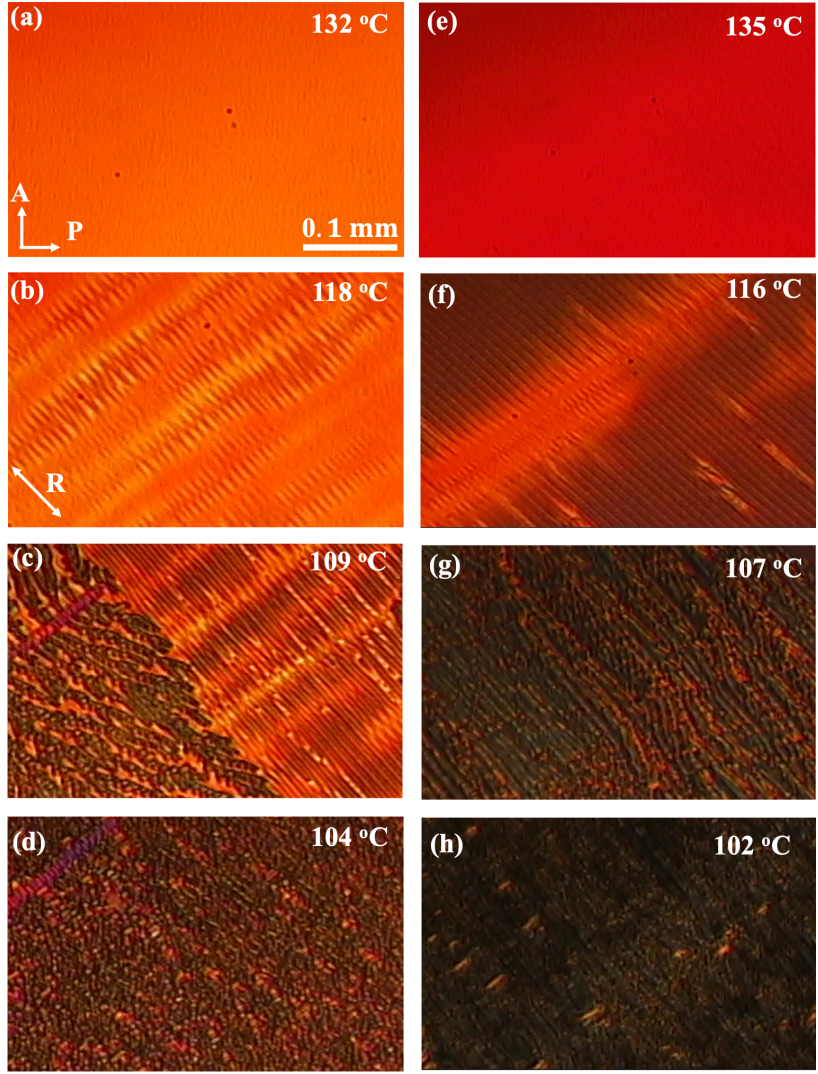

*Figure 2: Polarized optical microscopy (POM) textures of 5 μm, planar aligned **A** (left) & **B** (right) samples. Top row (a and e): nematic phase. Second row (b and f): $N_{TB}$ phase. Stripes are parallel to the rubbing direction (white double headed arrow in (b)). Third row (c and g): During and 0.5°C below $N_{TB}$-SmX transition, respectively. The textures are characterized by twisted ropes running parallel to the rubbing direction. Bottom row (d and h): Smectic textures about 5°C below the $N_{TB}$-SmX transition. The rope textures fade away.*

    C.    <u>*Small Angle X-ray Scattering (SAXS)*</u>

Small angle X-ray scattering (SAXS) measurements were carried out at beamline 7.3.3 of the Advanced Light Source (ALS) at Lawrence Berkeley National Laboratory with 10 KeV x-ray energy. Both samples of **A** and **B** were loaded in 2mm diameter quartz capillaries and placed in a customized Instec hot stage equipped with SmCo magnets, which produced a 1.5 Tesla magnetic field across the sample to align the nematic director. 2D SAXS patterns were recorded on a Pilatus 2M detector (Dectris, Inc.) at 2m distance from the sample. Each sample was heated to the isotropic phase and cooled slowly into the liquid crystal phases, where SAXS data were taken.

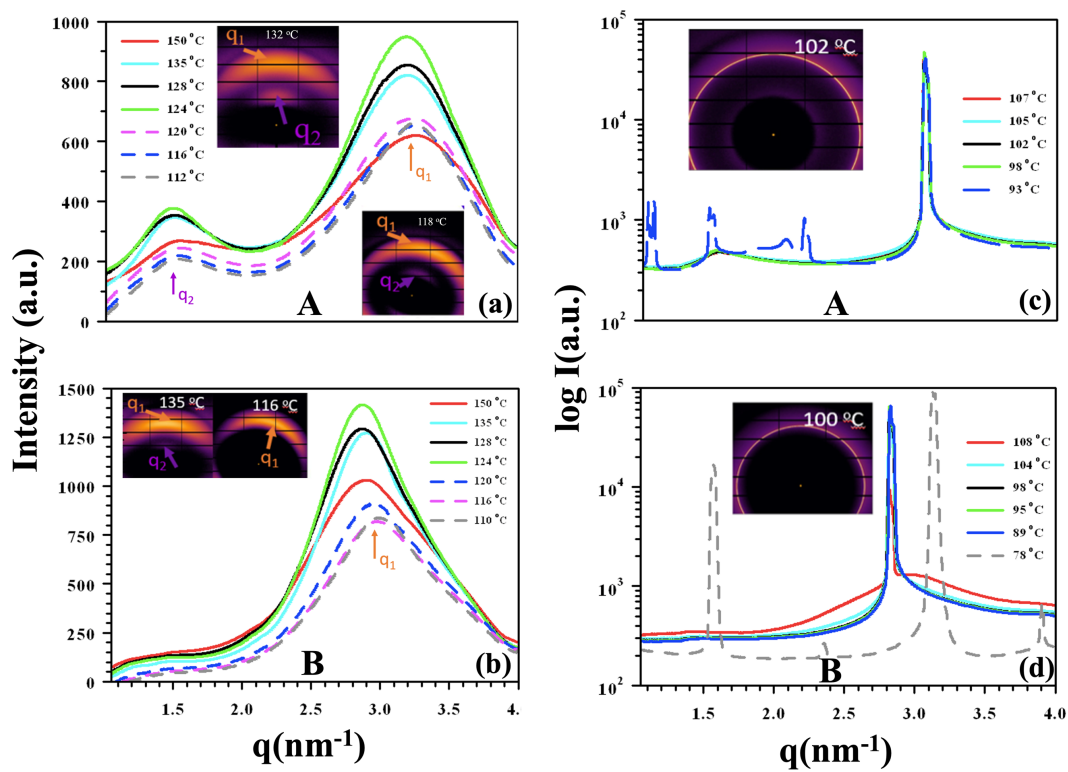

*Figure 3: Scattered intensity vs wavenumber (q) in $1<q<4$ nm$^{-1}$ range at selected temperatures of **A** (top: a,c) and **B** (bottom: b,d). Left side (a and b): q-dependence of the scattered intensity in the N (solid lines) and $N_{TB}$ (dashed lines) phases. Insets: 2D images at 132 °C and 118 °C of **A** (a) and at 135 °C and 116 °C of **B** (b). Dark lines are edges of the detector array. Right side (c and d): q-dependence of the scattering intensity in the SmA (solid lines) and crystal (dashed lines) phases. Insets: 2D patterns of SmA phase of **A** (c) and **B** (d).*

As seen in Figure 3(a), **A** exhibits two diffuse peaks both in the *N* (solid lines) and $N_{TB}$ (dashed lines) phases. These peaks are centered at $q_1 \sim 3.25$ nm$^{-1}$ and $q_2 \sim 1.53$ nm$^{-1}$, corresponding to $d_1 \sim 1.93$ nm and $d_2 \sim 4.10$ nm spatial periodicities. The intensity at $q_1$ is about 4 times larger than at $q_2$, representing more populated monomer-monomer (periodicity $d_1$, slightly smaller than half length of the dimer) than dimer-dimer (periodicity $d_2$ corresponding to the dimer length) associations.[22] From the full width at half maxima (*FWHM)*, we estimate the correlation length of the monomer-monomer association to be $\xi=1/FWHM \sim 8\text{-}10$ nm, corresponding to 2-3 molecular length. For **B** (Figure 3(b)) with the longer spacer, the peak with $q_2$ is basically missing, indicating only monomer-monomer type associations with $q_1 \sim 2.88$ $nm^{-1}$ corresponding to $d_1 \sim 2.1$ nm periodicity. This is likely due to the longer spacer that affords greater conformational freedom, suppressing the dimer-dimer correlations. Figures 3(c) and (d) show that on cooling below 108°C for **A** and 107°C for **B** the peaks at $q_1$ sharpen considerably, and the intensity increases by over three orders of magnitude indicating the formation of a smectic phase with layer periodicity being approximately half of the molecular length. For **A** only, there is an additional diffuse peak at $q_2$ (because this peak is so week, it is only discernible in a semi-logarithmic plot). This additional peak indicates some axially polar dimer-dimer associations within the apolar arrangement of molecules with head-tail symmetry. The peak position of $q_1$ in both **A** and **B** is basically independent of the temperature in the smectic phase.

As Figure 4(a) shows, the temperature dependences of the periodicities corresponding to the short-range monomer-monomer associations in the $N$ and $N_{TB}$ phases and of the layer spacing in the smectic phase are similar in the two materials. In the nematic phase, the periodicity of the molecular associations increases on cooling, reaches a plateau about *10°C* above the N-N$_{TB}$ transition, and then decreases (especially for **A** with shorter spacer) before reaching the transition. Such pretransitional behavior has been observed for a number of dimers[41–43] and can be understood as a tilt of the molecular axis in fluctuating $N_{TB}$ domains with respect to the nematic director.[44] The decrease of the periodicity continues below the $N$-$N_{TB}$ transition, but more weakly in **A**, and then starts to increase about *8°C* above the transition to the smectic phase. In case of **B**, the period of short-range monomer-monomer associations decreases from 2.19 nm to 2.08 nm, suggesting a tilt of $\theta \approx \cos^{-1}(2.08/ 2.19) \sim 17°$. This value is very similar to that found in the non-sulfur containing analogue, DTC5C9, as determined by the ratios of the smallest and largest periodicities of the cybotactic layer spacing measured above and below the N-N$_{TB}$ transition, respectively.[22] However, this apparent tilt is much smaller than that obtained using the ratio of the measured cybotactic layer spacing and the helical contour length between the centers of two mesogens, which yielded $\theta=29°$ for the DTC analogues.[30]

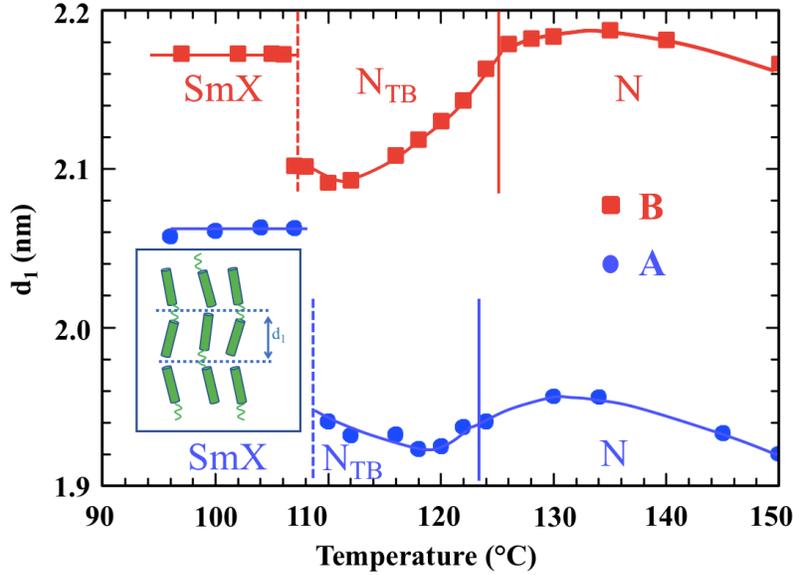

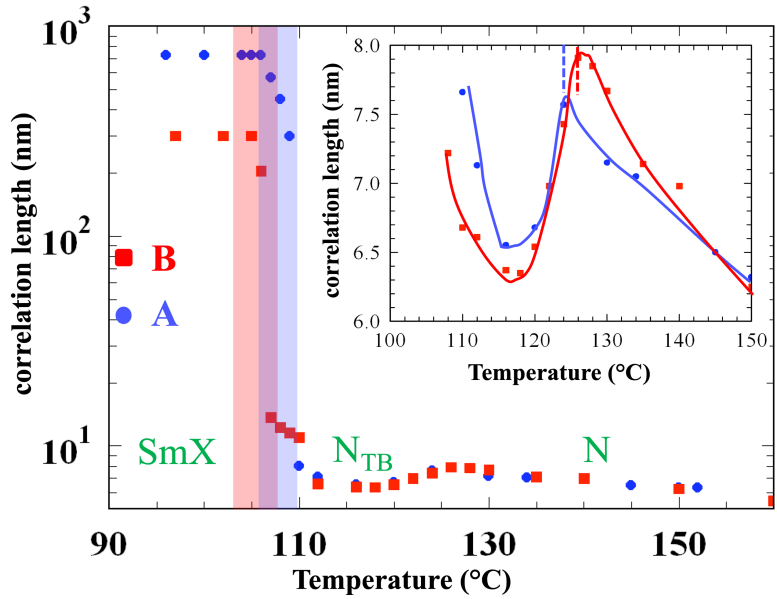

*Figure 4: Temperature dependences of the periodicity $d_1$ and the positional correlation length of **A** (blue dots) and **B** (red squares). (a): The periodicity calculated from the peak positions of the $q_1$ peaks. Box shows the suggested molecular packing in the smectic X phase; (b) Correlation length in logarithmic scale. Inset shows the correlation length in linear scale only in the N and $N_{TB}$ phases. Lines are guides to eye. Light blue and red highlights indicate the ranges for **A** and **B** where the rope-like texture resembling to the $N_{TB}$ texture was observable in POM studies (see Figure 2d and g).*

In the smectic X phase the layer spacings are *2.06 nm* and *2.17 nm* in **A** and **B**, respectively. These values correspond to roughly half of the dimer lengths, indicating apolar and random positions of the flexible spacers in neighbor molecules in a layer, as depicted in the inset to Figure 4(a). The difference in the layer spacings for **A** and **B** is consistent with the different flexible spacer lengths. Interestingly, for **A** the layer spacing is significantly higher than the lengthwise monomer-monomer separation in the N and $N_{TB}$ phases, whereas in **B** they are much closer.

As seen in Figure 4(b), for both materials the correlation length decreases upon the N-$N_{TB}$ transition from *7.6 nm* to *6.6 nm* for **A** and *8.0 nm* to *6.3 nm* for **B**, indicating a decrease in the short-range positional order. This is the same trend observed for the *DTCnCm* materials without sulfur atom.[22] Due to the underlying smectic phase, the correlation length increases gradually in the lower temperature range of the $N_{TB}$ phase, then grows sharply over a ~4°C range (shown by light blue and red highlights in Fig. 4(b)) to over 700 nm and 300 nm in the smectic phase of **A** and **B**, respectively.

Our SAXS results provide useful information on the orientational and positional order. However, they do not give any information about the heliconical pitch in the $N_{TB}$ phase; the latter may be determined from Tender Resonant X-ray Scattering.

D.   *Tender Resonant X-ray Scattering (TReXS)*

*TReXS* measurements were carried out at beamline 5.3.1 in the Advanced Light Source (ALS), Lawrence Berkeley National Laboratory. **A** and **B** were melt-loaded in between two silicon nitride membranes. Birefringence color indicated sample thicknesses in the range of 5-10μm. The samples were attached to a home-made heater which were sealed inside a helium chamber on the beamline. All samples were initially annealed in the isotropic state to remove heat history and defects. The X-ray beam energy was set at the sulfur K-edge by a channel cut double-bounce silicon monochromator.[32] A 2D detector (Pilatus 300K, Dectris, Inc) was used to collect the scattering patterns at a sample-detector distance of 393 mm. The beam center and the sample-to-detector distance were calibrated using silver behenate and the smectic A phase of 8CB. All TReXS data presented are measured on cooling at *1°C* rate after the samples heated to the isotropic phase. No well-defined features (rings/peaks) are seen in the scattering from the nematic phase. Intensity vs scattering wavenumber *(q)* curves for **A** and **B** were obtained from the 2D scattering patterns shown in the inset of Figure 5 using the Nika software package.[45]

The heliconical pitch (*p*) was calculated from the peak positions as *p=2π/q*. The temperature dependence of the pitch for both materials are plotted in Figure 5. After initially decreasing rapidly below the N-$N_{TB}$ phase transition, $p$ approaches an asymptotic value far from the transition, as observed for other dimers by RSoXS[27–30] and TReXS[32–36] techniques. Interestingly the temperature ranges where the nanoscale pitch was detected are 16.5°C and 22°C for **A** and **B**, respectively. These values are 3.3°C and 4.4°C larger than what we found by POM and SAXS measurements. As the N-$N_{TB}$ transition temperatures measured by the different techniques are the same within a degree, the difference is due to the fact that the nanoscale pitch disappears only at 103°C and 106°C, i.e., about 3.5° and 4.5°C below the transition to the SmX phase observed by SAXS. This strongly indicates that the nanoscale pitch survives in the top 3-4°C range of the smectic phase. The smectic layer periodicity of ~2 nm is not detected in our TReXS measurements, because they fall outside the range of our detector. After further cooling to the crystal phase, new peaks appear, corresponding to spatial periodicities of approximately 4, 6 and 8 nm indicating that they are harmonics of the 2 nm periodicity of the monomers.

The temperature dependence of pitch data measured in the $N_{TB}$ phase can be fitted to the expression

$$p(T) = p_o + \Delta p \cdot (1 - T/T_C)^{-\gamma} \qquad (1)$$

Here $p_o$ is the asymptotic pitch value very far from the critical temperature $T_C$, which is slightly (~1°C) larger than the *N-$N_{TB}$* phase transition temperature $T_{TB}$. The parameter $\Delta p$ is the coefficient of the temperature dependent term, and $\gamma$ is the critical exponent of the temperature dependent term. The four parameters in Eq. 1 are not sufficiently independent to reliably determine them by least squares fitting; indeed, it is possible to obtain reasonable looking fits with $\gamma$ ranging from 0.2 to 1. Motivated by predictions of macroscopic mean-field theories [23,46,47], we elected to fix the value of $\gamma$ to 0.5. Figure 5 shows fits to Eq. (1) for this fixed value. The fits give $p_o$ = *6.76 nm* and *6.35 nm, $\Delta p$ = 0.50 nm* and *1.07 nm* and $T_C$=*124.4°C* and *127.4°C* for **A** and **B**, respectively. These values are about 2°C higher than what we observed by POM measurements, where the appearance of the micrometer-range stripes was considered to be sign of the phase transition. Since the optical stripes are due to the Helfrich-Hurault-type instability[41], they may appear 1-2°C below the actual transition. Alternatively, they may indicate that the $\gamma$=0.5 exponent predicted by mean-field theory is not exactly accurate. The fact that $p_o$ is smaller for the longer molecule **B** than for **A** suggests that the second term in Eq. (1), expected only to describe the pretransitional temperature dependence of $p$, does not hold far from the transition. The ratios of the values of $\Delta p$ for **B** and **A** are approximately the same as the ratio of the corresponding changes in the spatial correlation lengths (see inset of Figure 4(b)).

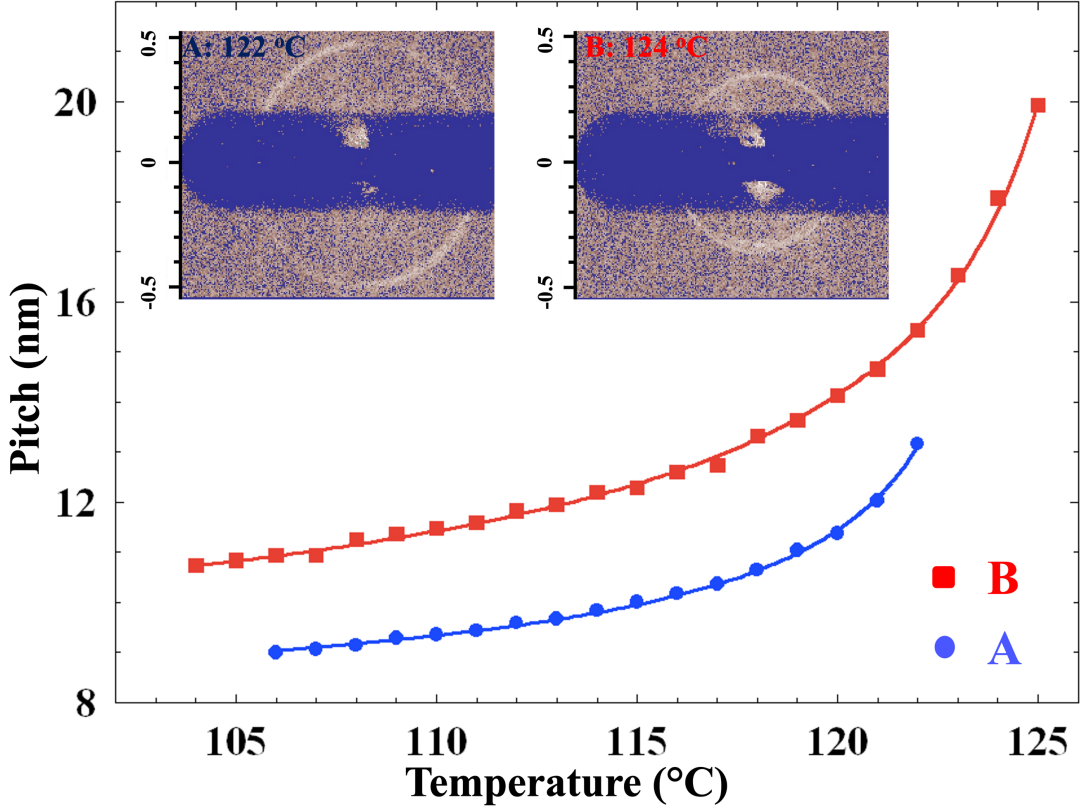

*Figure 5: Temperature dependences of the heliconical pitch of A (blue dots) and B (red squares). Solid lines with matching colors correspond to fit equation given in Eq. (1) with $\gamma=0.5$ and best fit parameters $p_o=6.76$ nm and 6.35 nm and $\Delta p=0.88$ nm and 1.88 nm for A and B, respectively. Pictures show the 2-dimensional scattering profiles for A and B at 122°C and 124°C, respectively. Scale numbers of the pictures are in $nm^{-1}$ unit.*

The temperature dependence of the correlation length of the heliconical bond order for **A** and **B** is shown in Figure 6 (a). The correlation length was calculated from the full width at half maxima (FWHM) of the peaks as $\xi = 1/\text{FWHM}$ (see inset of Figure 6(a)). The FWHM was determined by fitting the data for the scattered intensity vs $q$ to a Gaussian, $I(q) = a+b\cdot\exp(-(q - q_o)^2/\Delta q^2)$, and then taking FWHM$= 2\Delta q\,(\ln 2)^{1/2}$. The values of $\xi$ observed near the transition at $T = T_{TB}$ are only slightly larger than of the measured pitch at $T_{TB}$, corresponding to the situation where the heliconical structure just barely can form.

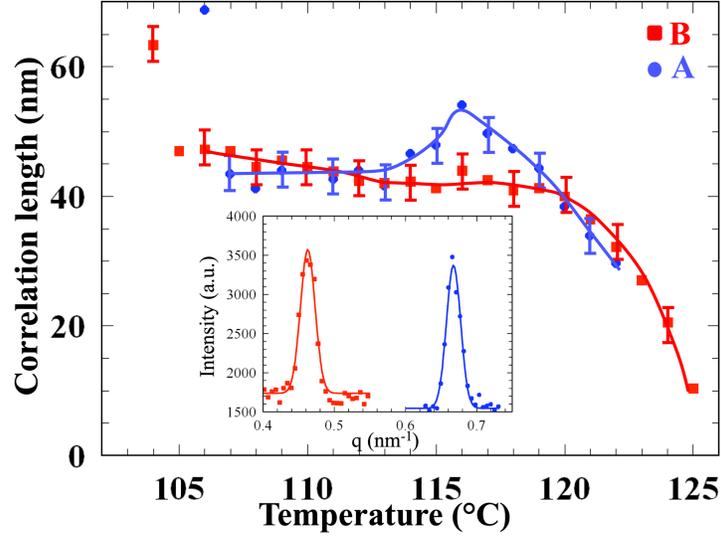

(a)

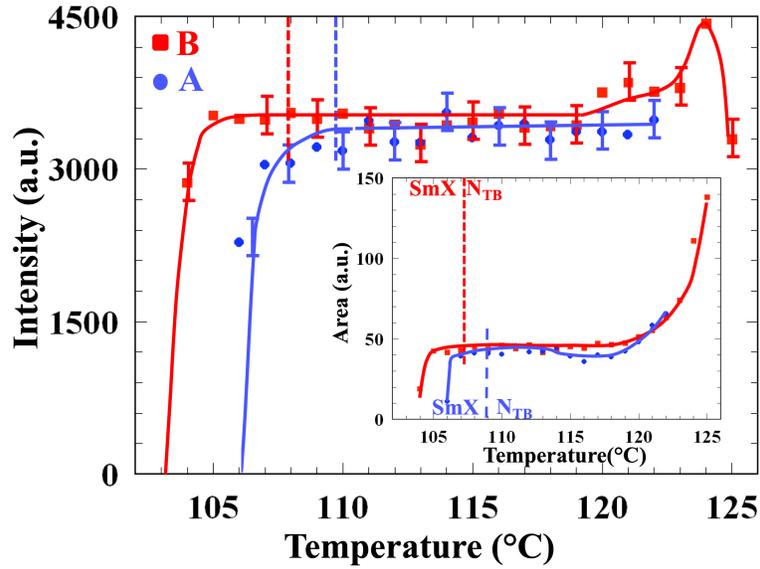

(b)

*Figure 6: Temperature dependences of the correlation length of the heliconical bond order (a) and the peak intensities (b) for **A** and **B**. Inset of (a) shows representative intensity vs q graphs with Gaussian fit to the measured data. Lines are guidance for the eye. Inset of (b) shows the areas below the peaks.*

For both materials $\xi$ increases sharply below $T_{TB}$ to *55 nm* and *44 nm* for **A** and **B**, respectively, which are 4-5 times longer than the heliconical pitch. The increase stops for **A** at around 116°C, then the correlation length decreases continuously to about 45 nm. For **B** there is a weak local minimum at 114°C, then the correlation length increases to about 48 nm. For both materials the peak width decreases (i.e., the apparent correlation length increases) as the peak height drops before the heliconical structure disappears at 106°C for **A** and 103°C for **B**.

The temperature dependences of the maximum intensities of the peaks and the area below the peaks are shown in the main pane and in the inset of Figure 6(b). For both materials the intensity is almost constant over the entire range except for a small maximum at 115°C for **A** and at 124°C for **B**. Importantly the intensity does not decrease in the last 3-4°C where the smectic phase has already formed, but only within 1°C before the peak disappears. Note, these correlation lengths are an order of magnitude smaller than those estimated for the prototypical cyano-biphenyl type dimer *CB7CB* from FFTEM measurements.[6,26]

## *3. Discussion*

The key observation of our experiments is that on cooling the dimers through the *N-N$_{TB}$* phase transition the correlation length of the spatial periodicity drops, while the heliconical orientational order becomes more correlated.

Formation of the *N$_{TB}$* phase, with short-range positional order and longer-range heliconical orientation (bond) order, reflects a subtle interplay between the molecular bend and flexibility. The twist-bend nematic phase forms as the odd-membered linkage becomes sufficiently rigid along one axis to provide bend with angle $\beta$ between the rigid arms and a twist with angle $\alpha$ between the planes of the arms (propeller type structure)[48]. Such molecules can only fill the space when twist and bend deformations couple, leading to the substantial bond order correlations we have observed. The formation of the heliconical bond order locks the molecular twist-bend conformation, i.e., it makes the linkage effectively rigid. This suppresses the entropic penalty of packing the flexible linkage next to a rigid arm, therefore favoring positional disorder as was found in this work. This means that the heliconical orientation is determined by the twist-bend molecular conformation, and does not require the positional correlations that would arise by "chaining" molecules to form a quasi-polymeric structure.[17] Therefore, the mechanism for the tight pitch of the twist-bend nematic liquid crystal phase (at least in these materials studied here) is very different from that of other systems such as proteins and DNA, which require both molecular chirality and more significant positional correlations that are generally absent in N$_{TB}$-forming dimers.

The fact that the correlation length of monomer-monomer aggregates drops below the transition to $N_{TB}$, can be interpreted as the result of an increase in the average molecular overlap between two molecules along the helical axis that we can define an overlap parameter as $J=l/L$, where $l$ is the length of molecular overlap and $L$ is the contour length of the dimers. For example, if the length of the rigid arms is about the same as of the connecting spacer, then in a cluster with associated monomer segments, the overlap is $J \sim 1/3$. For completely random positional order (zero spatial correlation length), however, $<J>=1/2$. An increased average overlap between the molecules would translate to a larger spatial periodicity assuming the same tilt of the molecules with respect to the cybotactic smectic layers. Alternatively, similar spatial periodicities can be obtained with larger overlap but larger tilt, as schematically illustrated in Figure 7(a). An increased average overlap therefore can explain the discrepancy between the $\theta \sim 17°$ apparent cone angle calculated from the ratio of the lowest and highest values of the periodicity of the layer spacing of the cybotactic clusters, and the ratio between the actual cybotactic layer spacing and the helical contour length between the centers of two mesogens, which provided $\theta = 29°$ for the DTC analogues[30] (see Figure 4(a) and the text above).

There have been several theoretical descriptions to describe the temperature dependence of the heliconical pitch. Models using continuum theory[23,46] predict that the pitch diverges as $p \propto (T_C - T)^{-1/2}$, where the difference between the critical temperature $T_C$ and the $N_{TB}$ phase transition temperature $T_{TB}$ ($T_C$-$T_{TB}$ ~ 1°C) is related to the molecular bend and flexoelectric coupling.[47] An alternative theoretical approach is based on microscopic molecular parameters, such as the bend and twist angles ($\beta$ and $\alpha$) between the molecular arms, and on pair-pair correlations.[49,48] This second approach also predicts temperature dependence qualitatively consistent with our measurements, but it is difficult to compare them quantitatively with our experiments. An elegantly simple microscopic packing model testable by experiments was proposed recently by Tuchband et al[28] and motivated by the measured temperature dependences found in *CB7CB*. It is based on the notion that, due to the molecular bend, the molecules could be packed to form a circle with perimeter $P=2\pi R_{mol}=k_0 L$, where $R_{mol}$ is the curvature radius of the bent molecules, $L$ is the contour length of the dimer, and $k_0$ is the number of molecules making one full circle, assuming zero overlap between neighbor molecules. Due to the effective twist between the molecular arms, and because circles with pure bend cannot fill the space effectively, the bend is coupled with twist and the molecules form a helicone with cone angle $\theta$ and pitch $p$ that are determined by the molecular twist and bend – the so-called "slinky" model. This model relates the pitch to the cone angle as

$$p(T) = 2\pi R_{mol} \cos\theta = k_0 \cdot L \cdot \cos\theta \qquad (2)$$

Using this equation, and even assuming the larger value of the tilt angle ($\theta=29°$) that was measured for the DTC analogs [30], we find the variation of the pitch for **B** between 125°C and 118°C should be less than *2.6 nm*, in contrast to the measured *9 nm* (see Figure 5). A similar discrepancy was noticed by Cruickshank et al[35] for several cyano-terminated sulfur containing materials, and they proposed that the parameter $k_0$ increases with temperature toward the $N_{TB}$-N transition. Implicitly this means that the molecular bend angle $\beta$ is increases in heating as the $N_{TB}$-N transition is approaching. Indeed, due to the flexibility and large number of configurations of the linkage, one expects $\beta$ to decrease with increased spacer length and increase on cooling. In frame of the slinky model $R_{mol} = L/\beta_{TB}$ and $k_0 = 2\pi/\beta_{TB}$, where $\beta_{TB}$ is the molecular bend at $T_{TB}$. Since $\beta_{TB}$ is expected to decrease with increasing length of the spacer, this explains the larger pitch measured for **B** compared to **A**. The observation that $p(T_{TB})$ is about twice as large for **B** with *n=11* carbons in the flexible linker than for **A** with *n=7*, cannot be understood simply by the difference between contour lengths $L_B$ and $L_A$. Since $L_B$-$L_A$ is equal to the length of 4 carbon-carbon bonds (4·0.154nm·cos60° ≈ 0.31nm), it gives less than a 10% increase. The observed $p_B(T_{TB})/p_A(T_{TB}) \sim 1.7$ can be explained from Eq. (3) with $\beta_A(T_{TB})/\beta_B(T_{TB}) \sim 1.5$.

Variation of molecular overlap with temperature should also affect the heliconical pitch. The number of molecules making one pitch with $J$ overlap ($0 < J < 1$) between two molecules is $k(T) = k_0(2J(T)+1)$. Combining this with $k(T) = 2\pi/\beta(T)$, we can express Eq. (2) as

$$p(T) = \frac{2\pi L \cos\theta(T)}{\beta(T)(2J(T)+1)}, \qquad (3)$$

Due to the longer spacing, the change of the overlap below the $N$-$N_{TB}$ phase transition is much more pronounced in **B** than in **A**. This can explain the observation that for **A** the smectic layer spacing is much higher than the periodicity of the monomer-monomer association, whereas in **B** they are about the same (Figure 4(a)).

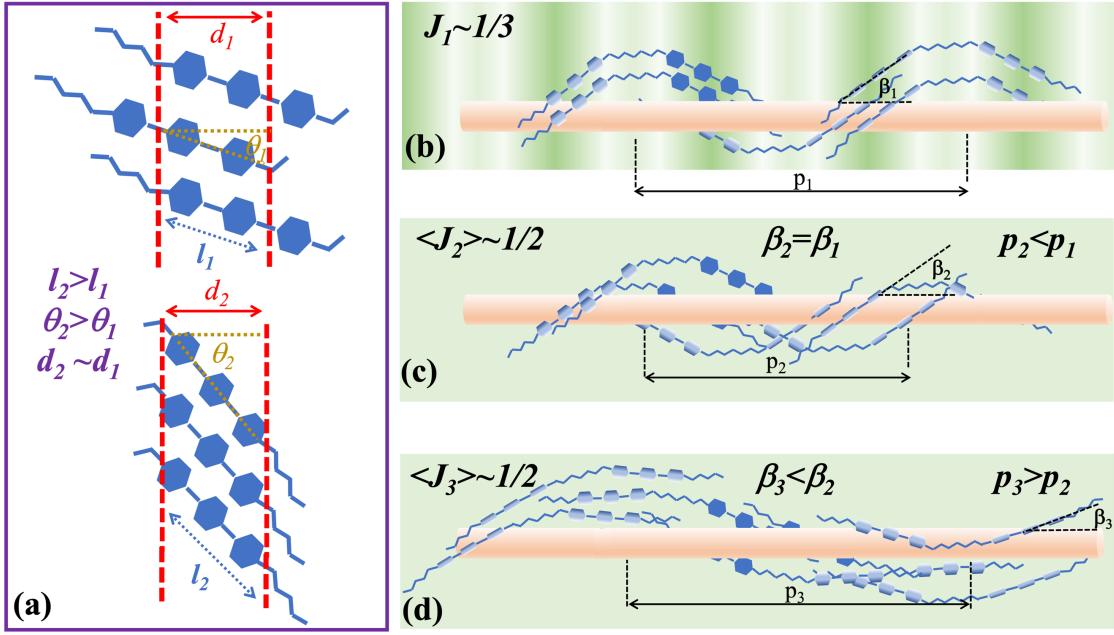

*Figure 7: Illustration of the heliconical structure of the $N_{TB}$ phase at various parameters. (a) Illustration of the interplay of overlap length l and the cone angle $\theta$ in determining the periodicity of monomer-type associations. (b) The heliconical structure with overlap parameter $J_1 \sim 1/3$. Variation of green shade in the background illustrates spatial periodicity. The shading illustrates short positional order in the lower range of the $N_{TB}$ phase and becomes long-range order upon the transition to the smectic phase. (c) Heliconical structure with random molecular positions along the heliconical axis that means the average overlap parameter $<J_2> \sim 1/2$. The molecular bend angle $\beta_2$ is the same as in (b) but the pitch is shorter. (d) Heliconical structure with the same average overlap parameter as in (c) ($<J_3> \sim 1/2$), but smaller molecular bend angle ($\beta_3 < \beta_2$) leading to pitch $p_3 > p_2$. Uniform green shade illustrates zero positional order.*

From Eq. (3) we also see that on cooling the pitch can decrease faster than $\cos\theta(T)$, if $\beta(T)$ or $J(T)$ (or both) increase on cooling. Thus, we may use Eq. (3) to explain the observed *9 nm* decrease of the pitch, instead of the 2.6 nm calculated from Eq. (2). The effects of $\beta$ and $J$ on the pitch are illustrated in Figure 7(b-d). Figure 7 (b) schematically shows the heliconical structure with overlap parameter $J_1 \sim 1/3$, molecular bend angle $\beta_1$ and pitch $p_1$. The variation of green shading in the background highlights the spatial periodicity associated with uniform overlap. The heliconical structure with random molecular positions along the heliconical axis (average overlap parameter $<J_2> \sim 1/2$) is shown in Figure 7(c). One can see that, while the molecular bend angle $\beta_2$ is the same as in Figure 7(b), the pitch is shorter. Figure 7(d) shows the heliconical structure with the same average overlap parameter as in Figure 7(c) ($<J_3> \sim 1/2$), but smaller molecular bend angle ($\beta_3 < \beta_2$), which leads to an increased pitch $p_3 > p_2$.

Another significant result of our experiments is the observed overlap between the smectic order and heliconical order in 3-4°C temperature range. The coexistence of the smectic order and the heliconical structure has been already reported by Abberly et al[50] as a new phase, the SmC$_{TB}$ phase, in dimers that did exhibited N and SmA phases but no N$_{TB}$ phase. In our case, the twist-bend type smectic phase is observed below the N$_{TB}$ phase and the heliconical structure disappears about 4°C below the transition without any observable DSC signals. SAXS measurements do not show any observable change in the layer periodicity. The intensity of the scattering peak and the spatial correlation length increase abruptly in this range and reach maxima before the heliconical structure disappears (see Figure 4(b)). At the same time, the TReXS peak intensities are almost constant and decrease abruptly only when the heliconical structure disappears. These indicate that both the smectic layers and the heliconical structure fill the entire volume simultaneously, thus forming a SmC$_{TB}$-type phase where the rigid units of the dimers are tilted with respect to the layer normal to allow for the bending of the dimers, but the tilt direction rotates following the heliconical order. Such a structure is similar to that shown in Figure 7(b) except that the spatial correlations are more pronounced. This structure can also explain the twisted rope POM texture that appear right below the N$_{TB}$-SmX transition with width and direction matching the stripe texture of the N$_{TB}$ phase. Future experimental and theoretical studies could determine if the smectic phase we observed in a 3-4°C range below the N$_{TB}$ phase is the same as the SmC$_{TB}$ phase found by Abberley et al below a SmA phase[50], and what is the mechanism of the disappearance of the heliconical order at lower temperatures.

## 4. *Conclusion*

To summarize, we have measured the nanostructure of two novel sulfur containing dimer materials both by hard Small Angle X Ray Scattering that is sensitive to positional order and by resonant tender X-ray scattering that can detect the heliconical bond order. Our most significant observations are the following: (a) On cooling the dimers through the *N-N$_{TB}$* phase transition the correlation length of the spatial periodicity drops, while the heliconical orientational order becomes more correlated. (b) The temperature dependences of the heliconical pitch show stronger variation near the N-N$_{TB}$ transition than in prototypical CBnCB-type dimers. (c) The heliconical pitch is observable even in the upper 3-4°C range of the underlying smectic phase.

The reduction in positional correlations upon the development of the heliconical order leads us to propose a temperature dependent variation of the molecular overlap. We discussed the temperature dependence of the heliconical pitch in terms of both a phenomenological mean field theory of the $N_{TB}$-N transition and in terms of a physical model for the molecular packing ("slinky" model). We found that the phenomenological theories qualitatively agree with our observations, while the "slinky" model predicts a weaker pitch variation than we measured. We proposed that a temperature dependent molecular bend and molecular overlap can account for the observed temperature variation and the spacer length dependence of the heliconical pitch.

The observed coexistence of smectic layering and heliconical order – both having periodicities on the scale of the molecular length – indicates a $SmC_{TB}$-type phase, where the rigid units of the dimers are tilted with respect to the layer normal to allow for the bent conformation of the dimers, but the tilt direction rotates along the heliconical axis

The results presented here indicate the value of employing multiple structural probes in order to illuminate the complex interplay between molecular shape, molecular flexibility, and intermolecular packing that governs the microscopic structure of liquid crystalline states, such as the twist-bend phase, that feature novel, nanoscale modulations of the molecular arrangements. The combination of orientational order, bond order, and short-range positional order in these systems represents a particular challenge to structural determination. Moreover, many theoretical models do not naturally lend themselves to experimental verification. By combining independent techniques that probe different correlations, we have demonstrated the possibility of determining with little ambiguity details about molecular arrangements that would otherwise be unavailable.

## 5. *Acknowledgement*

This research was supported by the National Science Foundation under grant DMR-1904167 and by (UK) EPSRC project EP/M015726/1. We acknowledge use of Beamlines 7.3.3 and 5.3.1 of the Advanced Light Source supported by the Director of the Office of Science, Office of Basic Energy Sciences, of the U.S. Department of Energy under contract no. DE-AC02-05CH11231.
C. Feng and R. Saha equally contributed to the paper.

(26) ... E.; Bedrov, D.; Walba, D. M.; Glaser, M. A.; Maclennan, J. E.; Clark, N. A. Chiral Heliconical Ground State of Nanoscale Pitch in a Nematic Liquid Crystal of Achiral Molecular Dimers. *Proc. Natl. Acad. Sci. U. S. A.* **2013**, *110* (40), 15931–15936. https://doi.org/10.1073/pnas.1314654110.

(27) Zhu, C.; Tuchband, M. R.; Young, A.; Shuai, M.; Scarbrough, A.; Walba, D. M.; Maclennan, J. E.; Wang, C.; Hexemer, A.; Clark, N. A. Resonant Carbon -Edge Soft X-Ray Scattering from Lattice-Free Heliconical Molecular Ordering: Soft Dilative Elasticity of the Twist-Bend Liquid Crystal Phase. *Phys. Rev. Lett.* **2016**, *116* (14), 147803. https://doi.org/10.1103/PhysRevLett.116.147803.

(28) Tuchband, M. R.; Shuai, M.; Graber, K. A.; Chen, D.; Zhu, C.; Radzihovsky, L.; Klittnick, A.; Foley, L.; Scarbrough, A.; Porada, J. H.; Moran, M.; Yelk, J.; Korblova, E.; Walba, D. M.; Hexemer, A.; Maclennan, J. E.; Matthew, A.; Clark, N. A. DOUBLE-HELICAL TILED CHAIN STRUCTURE OF THE TWIST-BEND LIQUID CRYSTAL PHASE IN CB7CB. *arXiv:1703.10787v1* **2017**.

(29) Salamończyk, M.; Vaupotič, N.; Pociecha, D.; Wang, C.; Zhu, C.; Gorecka, E. Structure of Nanoscale-Pitch Helical Phases: Blue Phase and Twist-Bend Nematic Phase Resolved by Resonant Soft X-Ray Scattering. *Soft Matter* **2017**, *13* (38), 6694–6699. https://doi.org/10.1039/C7SM00967D.

(30) Stevenson, W. D.; Ahmed, Z.; Zeng, X. B.; Welch, C.; Ungar, G.; Mehl, G. H. Molecular Organization in the Twist–Bend Nematic Phase by Resonant X-Ray Scattering at the Se K-Edge and by SAXS, WAXS and GIXRD. *Phys. Chem. Chem. Phys.* **2017**, *19* (21), 13449–13454. https://doi.org/10.1039/C7CP01404J.

(31) Henke, B. L.; Gullikson, E. M.; Davis, J. C. X-Ray Interactions: Photoabsorption, Scattering, Transmission and Reflection E = 50-30,000 EV, Z = 1-92. *At. Data Nucl. Data Tables* **1993**, *54* (2), 181–342. https://doi.org/10.1016/s0961-1290(05)71235-7.

(32) Cao, Y.; Feng, J.; Nallapaneni, A.; Arakawa, Y.; Zhao, K.; Liu, F.; Zhu, C. Elucidation of Distinct Molecular Resonance Effects in Twist Bend Nematic and Smectic A Liquid Crystals via Tender Resonant X-Ray Scattering. *arXiv* **2019**, 1907.11330v2.

(33) Salamończyk, M.; Mandle, R. J.; Makal, A.; Liebman-Peláez, A.; Feng, J.; Goodby, J. W.; Zhu, C. Double Helical Structure of the Twist-Bend Nematic Phase Investigated by Resonant X-Ray Scattering at the Carbon and Sulfur K-Edges. *Soft Matter* **2018**, *14* (48),
22